\documentclass{article}
\usepackage{ijcai25}

\usepackage{times}
\usepackage{soul}
\usepackage{url}
\usepackage{bm}
\usepackage[hidelinks]{hyperref}
\usepackage[utf8]{inputenc}
\usepackage[small]{caption}
\usepackage{graphicx}
\usepackage{amsmath}
\usepackage{amsthm}
\usepackage{booktabs}
\usepackage{algorithm}
\usepackage{algorithmic}
\usepackage[switch]{lineno}
\usepackage{amsmath, amssymb}
\usepackage{multirow}
\usepackage{color}
\DeclareUnicodeCharacter{FF1A}{:}

\title{
From General Relation Patterns to Task-Specific Decision-Making in Continual Multi-Agent Coordination
}

\author{
Chang Yao$^{1,2}$
\and
Youfang Lin$^{1,2}$\and
Shoucheng Song$^{1,2}$\and
Hao Wu$^{1,2}$\and
Yuqing Ma$^{3}$\\
Sheng Han$^{1,2}$\And
Kai Lv$^{1,2}$\thanks{Corresponding author: Kai Lv.}\\
\affiliations
$^1$School of Computer Science \& Technology, Beijing Jiaotong University, Beijing, China\\
$^2$Beijing Key Laboratory of Traffic Data Mining and Embodied Intelligence, Beijing, China\\
$^3$Institute of Artificial Intelligence, Beihang University, Beijing, China\\
\emails
\{yaochang,yflin,insis\_songsc,wuhao\_,shhan,lvkai\}@bjtu.edu.cn,
mayuqing@buaa.edu.cn
}

\begin{document}

\maketitle

\begin{abstract}

Continual Multi-Agent Reinforcement Learning (Co-MARL) requires agents to address catastrophic forgetting issues while learning new coordination policies with the dynamics team. In this paper, we delve into the core of Co-MARL, namely Relation Patterns, which refer to agents’ general understanding of interactions. 
In addition to generality, relation patterns exhibit task-specificity when mapped to different action spaces. To this end, we propose a novel method called General \textbf{R}elation \textbf{P}atterns-\textbf{G}uided Task-specific Decision-Maker (RPG). In RPG, agents extract relation patterns from dynamic observation spaces using a relation capturer. These task-agnostic relation patterns are then mapped to different action spaces via a task-specific decision-maker generated by a conditional hypernetwork. To combat forgetting, we further introduce regularization items on both the relation capturer and the conditional hypernetwork. Results on SMAC and LBF demonstrate that RPG effectively prevents catastrophic forgetting when learning new tasks and achieves zero-shot generalization to unseen tasks.

\end{abstract}

\section{Introduction}

Collaborative Multi-Agent Reinforcement Learning (MARL) \cite{oroojlooy2023CMARL} focuses on how multiple agents can coordinate to achieve global goals. Traditional research mainly concentrates on enhancing agents' cooperation abilities in stable environments \cite{rashid2020QMIX,yu2022MAPPO}.
However, in the real world, agents often face constant changes in diverse tasks \cite{kiran2021RL4AD,yu2021RC,zhang2024enhancing,wang2024effects,wang2024MIIR}. Particularly, in continual coordination scenarios, agents may lose the ability to perform well in previously learned tasks when adapting to new ones, a phenomenon known as catastrophic forgetting \cite{robins1995CF} in Continual Learning (CL). 

CL seeks to exploit both the shared and the specific properties across different tasks. The properties reflect the essence of continual learning across various domains. For instance, in the class-incremental learning domain, models should understand the overlap or complementarity between new and old classes in feature space, leveraging learned knowledge to adapt to new tasks \cite{ji2023CIL}. In the multimodal CL domain, models learn sample-specific and invariant features to construct representations that are broadly adaptable, to prevent catastrophic forgetting \cite{zhang2023vqacl}. Compared to class-incremental and multimodal CL, the challenges in Continual Multi-Agent Reinforcement Learning(Co-MARL) are more complex due to the uncertain scale of entities.
Thus, it is also crucial to figure out the essence of Co-MARL. In this paper, we argue that the essence of continual multi-agent coordination lies in \textbf{Relation Patterns} between tasks.

\begin{figure}
    \centering
    \includegraphics[width=1.\columnwidth]{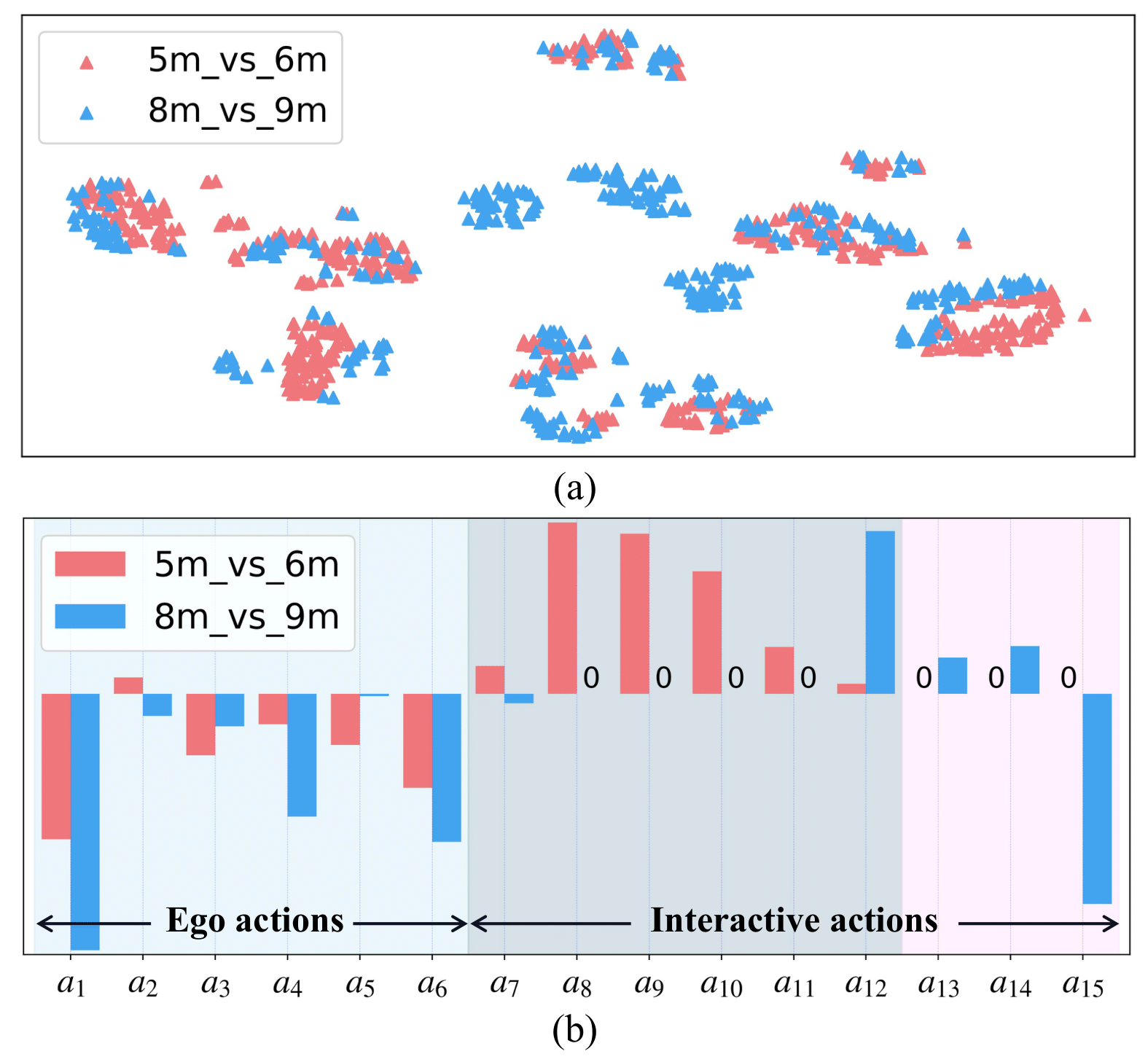}
    \caption{(a) Clustering visualization of relation patterns; 
    (b) Agent's action selection in the similarity relation patterns. The purple background represents the action space unique to the 8m\_vs\_9m.}
    \label{FIG:1}
\end{figure}

A \textbf{relation} represents the influence of an observable entity on an agent's decision-making. 
Specifically, the relation between the agent and each entity is determined by the combination of the entity's importance to the agent and the entity's features.
In this paper, a \textbf{relation pattern} is formed by integrating the relations of multiple entities.
Through experiments, we show that common relation patterns exist across distinct tasks. In Figure \ref{FIG:1}(a), we apply t-SNE \cite{van2008Tsne} to analyze relation patterns in two cooperative tasks. 
The red and blue triangles representing the two tasks are tightly clustered, indicating strong similarity in the relation patterns between the two tasks.
As the entity scale increases from task 5m\_vs\_6m to task 8m\_vs\_9m in SMAC \cite{samvelyan2019SMAC}, some blue triangles form additional isolated clusters.
This reflects that in a more complex observation space, agents can form more relation patterns.

Although the clustered relation patterns exhibit similarity across different tasks, the value of similar relation patterns on decision-making can vary significantly depending on the specific task context. For example, in the 8m\_vs\_9m, agents can win a game by capturing various ``focus-fire'' patterns (e.g., 4-on-2 or 3-on-2). 
However, in the 5m\_vs\_6m, the same ``focus-fire'' patterns may lead to a loss when applying the same policies.
This indicates that identical relation patterns result in different returns across tasks.
We validate this preference difference in Figure \ref{FIG:1}(b), where the agents demonstrate noticeable differences in their decision-making preferences (Q-value distribution) across two action spaces.

To \textbf{capture} and \textbf{exploit} these general relation patterns, we propose RPG(General \textbf{R}elation \textbf{P}atterns-\textbf{G}uided Task-specific Decision-Maker).
1) In \textbf{capturing} the relation patterns, we design a scalable relation capturer that allows agents to measure interactions with other entities from their first-person perspective, forming general relation patterns. We further add an anti-forgetting regularization to retain key relation patterns across tasks with varying entity numbers.
2) In \textbf{exploiting} the relation patterns, a task-specific decision-maker is introduced, enabling the agent to respond to dynamic threats and adapt to new collaborative teams.
We use a conditioned hypernetwork to generate decision-makers for each task, mapping similar relation patterns to diverse action spaces for flexible adaptation. To ensure the stability of each decision-maker, we add a hypernetwork loss term that prevents forgetting prior cooperation while enabling fast adaptation to new teams.
The contributions are as follows:
\begin{itemize}
    \item We capture the essence of Co-MARL via relation patterns, linking dynamic observation and action spaces.
    \item We introduce a scalable Relation Capturer that helps agents extract relation patterns, preventing catastrophic forgetting and enhancing cross-task collaboration.
    \item We inject plasticity into decision-makers through the hypernetwork, enabling flexible decisions based on relation patterns in dynamic environments.
    \item RPG outperforms existing baselines in continual tasks, excelling on benchmarks like SMAC\cite{samvelyan2019SMAC} and LBF\cite{papoudakis2020LBF}.
\end{itemize}

\section{Related Work}

\subsection{Continual Learning in RL and MARL}
Continual Reinforcement Learning (CRL) aims to enable agents to learn sequential tasks while avoiding catastrophic forgetting. 
Early approaches \cite{rolnick2019CLEAR,chaudhry2019Coreset} alleviate forgetting by reusing past experiences through replay buffers. Regularization-based methods \cite{kirkpatrick2017EWC,aljundi2018MAS} mitigate forgetting by constraining parameter updates to ensure stable learning. Another class of methods tackles the stability-plasticity trade-off by dynamically adjusting model architectures, such as employing new networks \cite{kim2023ANCL} or multi-head structures \cite{kessler2022OWL}. However, while these methods have shown progress in single-agent tasks, achieving effective continual learning in multi-agent tasks remains a significant challenge.

In Co-MARL, the complex interactions between agents make continual learning significantly more challenging. Agents must coordinate in diverse teams, requiring them to acquire new knowledge while retaining the ability to address previous tasks. To address these challenges, MALT \cite{shi2021MALT} employs progressive neural networks to facilitate knowledge transfer across tasks, while MACPro \cite{yuan2024MACPro} leverages a shared feature extractor and dynamically expands policy heads to effectively prevent catastrophic forgetting. LL-Hanabi \cite{nekoei2021LL-Hanabi} provides a benchmark for testing continual learning in multi-agent settings. MARR \cite{yangMARR} adopts a reset strategy to avoid the loss of network plasticity caused by high replay rates. However, these methods fail to thoroughly analyze the fundamental bottlenecks of Co-MARL. In this paper, our approach tackles the core challenges of Co-MARL, providing a stable and efficient solution to address catastrophic forgetting.

\subsection{Hypernetworks in MARL}
Hypernetworks have demonstrated strong performance in meta-reinforcement learning and continual learning \cite{beck2023hypermeta1,chandra2023hypercl1,ehret2020hypercl2}. In MARL, QMIX \cite{rashid2020QMIX} introduces hypernetworks into the mixing network to enable effective credit assignment, while MAVEN \cite{mahajan2019maven} is the first to leverage hypernetworks to generate decision-layer parameters, facilitating efficient exploration for agents. HPN \cite{jianye2022HPN} employs hypernetworks to achieve permutation equivariance and invariance, enabling it to handle observations with varying dimensions. Concord \cite{guanoneConcord} pioneers the use of hypernetworks for continual AI-human collaboration by generating agent parameters to adapt to diverse teammate strategies, though it incurs significant computational overhead and overlooks the challenge of varying agent numbers in continual tasks. In this work, we utilize hypernetworks to generate decision parameters for agents, providing the capability to learn new tasks and achieve stable continual coordination.

\section{Problem Formalization}

In a sequence of continual multi-agent coordination tasks, all agents can only access partial local observations. We model this setting as a Continual Decentralized Partially Observable Markov Decision Process (Co-Dec-POMDP) \cite{oliehoek2016Dec-POMDP}. The Co-Dec-POMDP for each task is formalized as a tuple:
\[
\mathcal{Z}_c = \langle c, \mathcal{N}_c, \mathcal{M}_c, \mathcal{S}_c, \mathcal{O}_c, \mathcal{A}_c, \mathcal{R}_c, \mathcal{P}_c, \gamma \rangle,
\]
where, in task \(c\), \(\mathcal{M}_c\) denotes the set of all entities in the environment, and \(\mathcal{N}_c \subseteq \mathcal{M}_c\) represents the set of agents. At each time step \(t\), the global state \(s_t \in \mathcal{S}_c\) represents the complete state of the environment. Each agent \(i \in \mathcal{N}_c\) receives a local observation \(o_t^i \in \mathcal{O}_c\) and selects an action \(a_t^i \in \mathcal{A}_c\). The joint action \( \boldsymbol{a}_t = \{ a_t^i \}_{i=1}^{|\mathcal{N}_c|} \) is then executed, and the environment transitions to the next state \(s_{t+1}\) according to the state transition function \( \mathcal{P}_c(s_{t+1} \mid s_t, \boldsymbol{a}_t) \), while generating a shared reward \(r_t \in \mathcal{R}_c\). The objective of the multi-agent system is to maximize the global value function for the current task: \(Q_{tot}(s, \boldsymbol{a}) = \mathbb{E}_{s_t, \boldsymbol{a}_t} \left[ \sum_{t=0}^{\infty} \gamma^t r(s_t, \boldsymbol{a}_t) \right]\), where \(\gamma \in [0, 1]\) is the discount factor, and \(\pi\) represents the joint policy.

In a Co-Dec-POMDP, when a new task \( c+1 \) arrives, the entity composition changes from \( \mathcal{M}_c \) to \( \mathcal{M}_{c+1} \), resulting in corresponding dimensional changes in \( \mathcal{S} \), \( \mathcal{O} \), and \( \mathcal{A} \). Within our continual learning framework, agents are designed to handle these dynamically changing spaces, with the objective of maximizing the global reward in the current task while maintaining the performance of the current agent network across all previously encountered tasks.

\section{Method}

\begin{figure*}
	\centering
	\includegraphics[width=1.0\textwidth]{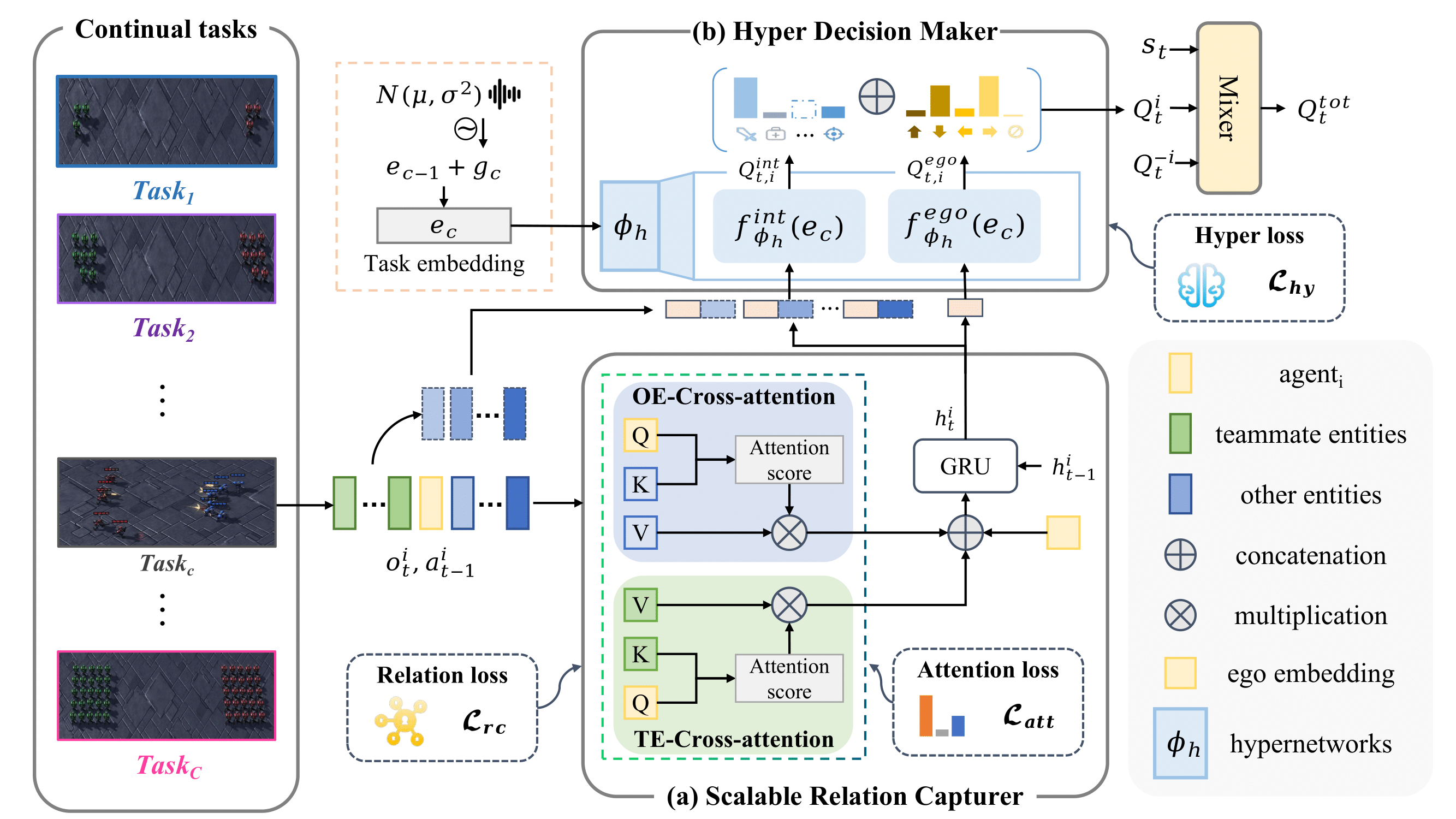}
	\caption{An overview of the RPG framework. In the \(c\)-th task, the observations are input into the \textbf{(a) Scalable Relation Capturer}, which captures the historical relational patterns. These captured patterns are then fed into the \textbf{(b) Hyper Decision-Maker}, where the hypernetworks generate the decision-makers that map the relation patterns to local Q-values.}
	\label{FIG:2}
\end{figure*}

We believe the core of Co-MARL lies in the relation patterns that manifest distinct properties across different spaces, i.e., observation and action spaces. On the one hand, general relation patterns emerge across tasks with varying numbers of entities, facilitated by the decomposition and combination of the observation space in an entity-wise manner. On the other hand, even similar relation patterns may yield different values when mapped to distinct action spaces.
To address this, we propose a General Relation Patterns-guided Task-specific Decision-Maker (RPG) framework. First, to capture and preserve the general agent-to-entity relations from the dynamic observation space, we design a scalable relation capturer with stability. Next, we inject plasticity into the decision-maker using a hypernetwork, enabling it to map the general relation patterns to task-specific action spaces.

\subsection{Scalable Relation Capturer with Stability}

In sequential Co-MARL tasks, extracting general relation patterns and maintaining their stability is crucial. To this end, we develop a scalable relation capturer to dynamically capture the interaction relationships between entities. The scalable relation capturer has two components: attention-based relation patterns and return-aware anti-forgetting.

\paragraph{Attention-Based Relation Patterns.}
To extract relation patterns at the entity level, we model the observation with non-fixed dimensions as a combination of multiple entity features. Consider a multi-agent system with \( m \) entities. Given the agent’s observation \( o_{t,i} \) at time step \( t \), we decompose \( o_{t,i} \) into the agent’s own features \( o_{t,i}^\text{se} \), \( n-1 \) teammate entity features \( o_{t,i}^\text{te} \), and \( m-n \) other entity features \( o_{t,i}^\text{oe} \).

Since the relation patterns are independent of the position of entities in the sequence, a permutation-invariant cross-attention mechanism can be used to capture general relation patterns.
Specifically, as shown in Figure \ref{FIG:2}(a), we treat \( o_{t,i}^\text{se} \) as the query, while \( o_{t,i}^\text{te} \) or \( o_{t,i}^\text{oe} \) serve as the keys and values, respectively. We then calculate the attention scores for teammate entities and other entities, corresponding to collaborative partners and potential threats or targets, respectively.
\begin{equation}
{\alpha}_{t,i}^\text{te} = \text{Softmax}(\frac{q K_\text{te}^\top}{\sqrt{d_x}}), {\alpha}_{t,i}^\text{oe} = \text{Softmax}(\frac{q K_\text{oe}^\top}{\sqrt{d_x}}),
\end{equation}
where \( \alpha^{\cdot}_{t,i} \) represents the attention level of agent \( i \) towards teammates and other entities , and \( \sqrt{d_x} \) is the scaling coefficient. Additionally, we use \( \alpha^{\cdot}_{t, i} \) to perform a weighted sum of values, thereby obtaining the relation pattern from the agent's first-person perspective to teammates and others, as shown below:
\begin{align}
z^\text{te}_{t,i} = {\alpha}^\text{te}_{t,i} V_\text{te}, z^\text{oe}_{t,i} = {\alpha}^\text{oe}_{t,i} V_\text{oe}.
\end{align}

The relation pattern \( z^{\cdot}_{t, i} \) is formed by concatenating the outputs from two cross-attention modules. Furthermore, to establish closer relation patterns with more relevant entities, inspired by MAIC \cite{yuan2022MAIC} and CoDe \cite{song2025code}, we introduce a regularization term to sparsify the attention scores:
\begin{align}
\mathcal{L}_\text{att} = \frac{1}{T} \sum_{t=1}^{T} \mathcal{H}(\alpha^{\cdot}_{t}) = -\frac{1}{T} \sum_{t=1}^{T} \sum_{i} \left( \alpha^{\cdot}_{t,i} \log \alpha^{\cdot}_{t,i} \right) .
\label{entro}
\end{align}

By minimizing the entropy of the attention scores \( \alpha_i^{\cdot} \), we encourage the agent to focus on several key entities. 

Finally, to capture the agent’s information, we encode its own features \( o_{t, i}^\text{se} \) as \( z_{t, i}^\text{se} \), which is then concatenated with the relation pattern to form \( z_{t, i} = [z_{t, i}^\text{se}, z_{t, i}^\text{te}, z_{t, i}^\text{oe}] \). It is then fed into the GRU along with the previous hidden state to generate the historical relation pattern \( h_{t, i} = \text{GRU}(z_{t, i}, h_{t-1, i}) \), 
which serves as the output of the scalable relation capturer.

\paragraph{Return-Aware Anti-Forgetting.}
Although we can extract common relation patterns across different tasks, excessive parameter updates of the capturer can lead to the forgetting of past relation patterns. Therefore, applying varying degrees of constraints to the different parameters ensures the stability of the capturer. Inspired by Taylor Pruning \cite{molchanov2019Taylorp}, under the i.i.d. assumption of parameters, we measure the importance of a parameter \( \theta_i \) by the squared difference in the TD loss when \( \theta_i \) is set to zero:
\begin{equation}
\mathcal{L}_{TD} = \mathbb{E} \left[ ( Q_{tot}(s, \mathbf{a}) - ( r + \gamma \max_{\mathbf{a}'} Q_{tot}(s', \mathbf{a}') ) )^2 \right] \label{TDloss},
\end{equation}

\begin{equation}
I(\theta_j) = (\mathcal{L}_{TD}-\mathcal{L}_{TD}(\theta_j=0))^2 \label{import}.
\end{equation}

Directly computing the parameter importance using Equation~\eqref{import} is computationally expensive. Therefore, we approximate the importance of \( \theta_j^c \) for the $c$-th task using a Taylor expansion. The second-order Taylor expansion is as follows:
\begin{equation}
I(\theta_j^c) = \left( \left( \frac{\partial \mathcal{L}_{TD}}{\partial \theta_j^c} \right) \theta_j^c + \frac{1}{2} \left( \frac{\partial^2 \mathcal{L}_{TD}}{\partial (\theta_j^c)^2} \right) (\theta_j^c)^2 \right)^2 .
\label{2taylor}
\end{equation}

We typically retain only the quadratic term in Equation \eqref{2taylor}. The final average importance of \( \theta^{c} \) is then given by:
\begin{equation}
I(\theta^{c}_j) = \frac{1}{N} \sum_{n'=1}^N \left( \left( \frac{\partial \mathcal{L}_{TD}^{c}}{\partial \theta^{c}_j} \right) \theta^{c}_j \right)^2 / I(\theta^{c})_\text{max},
\end{equation}
where \( N \) denotes the number of samples from the previous buffer, and all parameter importance values of \( \phi_\text{rc} \) are normalized by dividing by the maximum \( I(\theta^{c})_\text{max} \). 
Specifically, for the importance matrices \( I(\theta^1), I(\theta^2), \dots, I(\theta^{c-1}) \) of the past \( c-1 \) tasks, the parameter regularization item can be expressed as the sum of the regularization losses for the relation capturers of each task:
\begin{equation}
\mathcal{L}_\text{rc} = \frac{\lambda_\text{rc}}{2} \sum_{c'=1}^{c-1} \sum_{j=1}^{|\phi_\text{rc}|} I(\theta_j^{c'})(\theta^*_j - \theta_j^{c'})^2,
\end{equation}
where \( \lambda_\text{rc} \) represents the regularization strength for the relation capturer parameters \( \phi_\text{rc} \).
The learning process can be stabilized by penalizing inconsistencies in the parameters, which helps the agent retain key relation patterns and avoid catastrophic forgetting.

In addition, overly conservative regularization can prevent the agent from extracting all possible relation patterns in more complex tasks. During the early stages of learning a new task, the team's return reflects its ability to solve the current problem. 
Therefore, we use the team's normalized cumulative return over the first \( N \) episodes as the discount factor for \( \mathcal{L}_\text{rc} \):
\begin{equation}
\gamma_{p} = \frac{1}{N} \sum_{n'=1}^{N} G^{n'}_\text{norm},
\end{equation}
where $G^{n'}_\text{norm}$ denotes the raw cumulative return of the $n'$-th episode.
A larger $\gamma_p$ indicates that the current policy is well-suited to the current task, while a smaller $\gamma_p$ suggests relaxing the constraints in the new observation space to learn more potential coordination forms.

\subsection{Hyper-Decision-Maker with Plasticity}

Another key challenge in Co-MARL is that general relation patterns exhibit varying values across continual tasks. Agents must learn a new decision-making policy to adapt to changing action spaces. Therefore, we introduce a task-conditioned hypernetwork, which aims to generate task-specific decision-makers that can make rapid behavioral adjustments based on the captured relation patterns.

Using a traditional MLP as the policy layer may lead to a loss of plasticity in the agent. Instead, we utilize a hypernetwork to generate the decision-maker, which further maps the historical relation pattern to the action space. When learning the first task, we use a randomly initialized, Gaussian-distributed, learnable task embedding \( e_1 \) as the input to the hypernetwork \( \phi_h \). 

The hypernetwork generates an ego-decision-maker (right) and an interaction-decision-maker (left), as shown in Figure \ref{FIG:2}(b), corresponding to fixed ego actions and variable interaction actions, respectively.
For ego actions, we use the hypernetwork \( \phi_h^\text{ego} \) to generate the ego-decision-maker \( f_{\phi_h}^\text{ego}(e_1) \), and then map the historical relation pattern \( h_{t,i} \) to a fixed-dimension \( Q_{t,i}(a_\text{ego}|o_{t,i}) \). 
For interactive actions, directly mapping \( h_{t, i} \) to variable-dimension \( Q_{t,i}(a_\text{int}|o_{t,i}) \) is not feasible.
Therefore, we design an entity-wise interaction decision-maker that focuses on entity interactions. By concatenating the historical relation pattern \( h_{t,i} \) with the features of \( m-n \) other entities \( e_{t,i}^\text{oe} \), we map them to each interaction entity's \( Q_{t,i}(a_\text{int}( \cdot)|o_{t,i}) \). Finally, we concatenate the state-action values from both parts to obtain the complete local state-action \( Q_{t, i}(a|o_{t, i}) \).

In addition, the hypernetwork itself faces the issue of catastrophic forgetting across tasks.
We introduce an additional regularization term for the hypernetworks:
\begin{equation}
\mathcal{L}_\text{hy} = \frac{\lambda_\text{hy}}{c-1} \sum_{c'=1}^{c-1} \left\| f_{\phi_h}(e_{c'}) - f_{\tilde{\phi}_h}(e_{c'}) \right\|_2^2,
\end{equation}
where \( \tilde{\phi}_h \) refers to the hypernetworks obtained after the \( c-1 \)-th task. By minimizing the above loss term, we encourage the hyper-decision-maker to learn new behaviors while preserving the decision-maker specific to previous tasks. Furthermore, to ensure a smooth task transition, we initialize the task embedding \( e_c \) for the new task as follows:

\begin{equation}
e_c = \alpha_\text{init} e_{c-1} + (1 - \alpha_\text{init}) g_{c},
\end{equation}
where \( \alpha_\text{init} \) controls the degree of retention of the previous task's embedding, and \( g_{c} \) is the embedding initialized from a Gaussian distribution. This soft initialization method allows the agent to inherit knowledge from previous tasks and smoothly transition to the new task.

\subsection{Overall Algorithm}
In this section, we provide an overall description of RPG and present the pseudocode of our algorithm in the Appendix.

In the training phase of each task, we train the overall agent network and mixing network using the loss terms in Equation~\eqref{entro} and ~\eqref{TDloss}. Additionally, We integrate a multi-head attention module into the mixing network to accommodate the varying state spaces across tasks. This design facilitates the agent's ability to learn collaborative behaviors from a global perspective. A detailed explanation of the mixing network is provided in the Appendix.

Furthermore, during the training phase of the \( c \)-th (\( c > 1 \)) task, to prevent catastrophic forgetting, we introduce regularization loss terms for the relation capturer and decision-maker. The final overall loss function is as follows:
\begin{equation}
\mathcal{L}^{c}_{tot} = \mathcal{L}_{TD} + \alpha_\text{att} \mathcal{L}_\text{att} + \mathbf{1}_{c>1} \gamma_{p} \mathcal{L}_\text{rc} +  \mathbf{1}_{c>1} \mathcal{L}_\text{hy},
\end{equation}
where \( \alpha_\text{att} \) is a coefficient used to balance the various loss terms. Notably, \( \mathbf{1}_{c>1} \) is an indicator function that takes the value 0 in the first task (\( c = 1 \)) and 1 otherwise. Additionally, as tasks are trained sequentially, we evaluate the current agent network on the prior \( c-1 \) tasks to assess the anti-forgetting effect of RPG.

\section{Experiments}

\begin{figure*}
	\centering
	\includegraphics[width=1.0\textwidth]{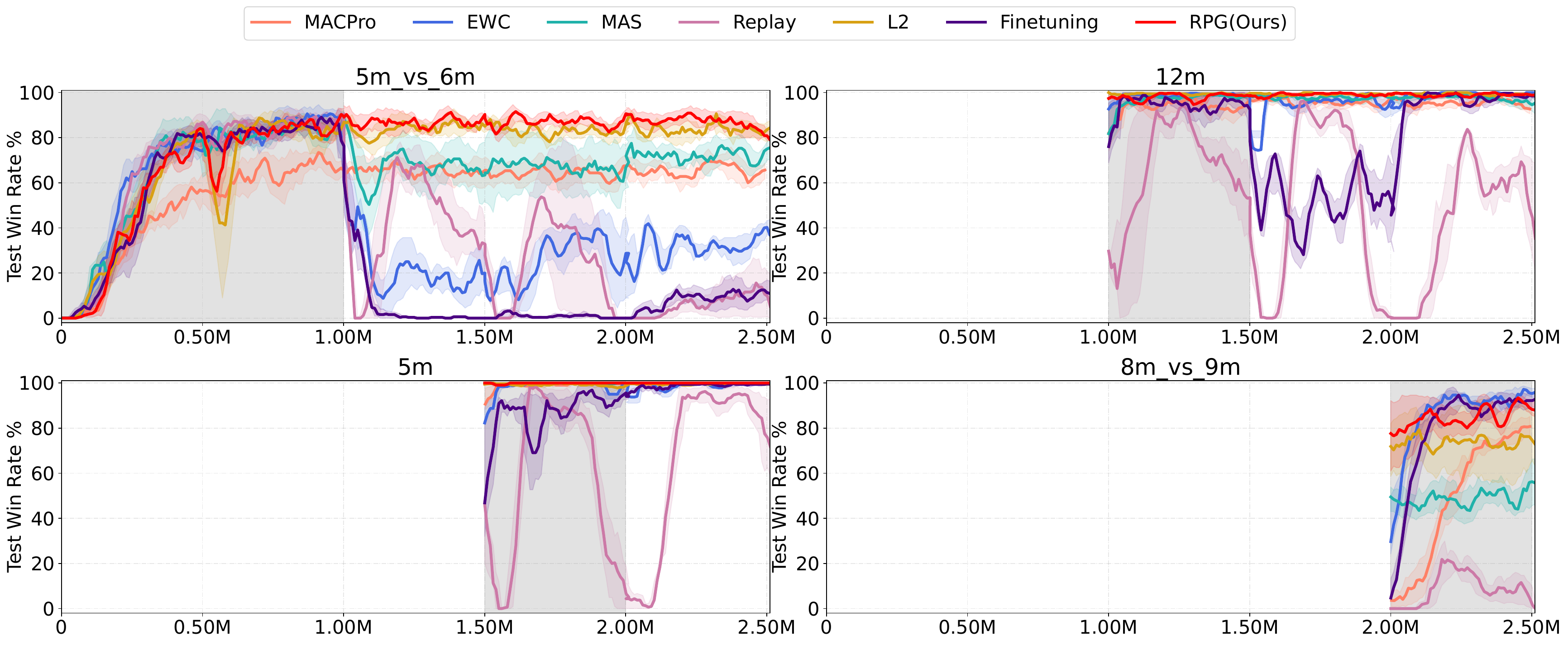}
	\caption{The complete continual learning results in StarCraft II. The tasks appear in the order of \{5m\_vs\_6m, 12m, 5m, 8m\_vs\_9m\}. The gray background indicates that the task is in the training phase, and the blank areas represent tasks that have not yet been encountered. The curves after the gray background represent the performance of the agent trained later when tested on that task. Each curve represents the average win rate across 3 random seeds.}
	\label{FIG:3}
\end{figure*}

To validate the effectiveness of RPG, we design comprehensive experiments to address the following questions:  1) Does our method outperform other baselines in learning new knowledge while preventing catastrophic forgetting? 2) How does each component of our approach impact continual learning performance? 3) Does the relation pattern extracted by RPG facilitate better multi-agent coordination compared to other baselines? 4) Does the general relation pattern captured by RPG contribute to zero-shot generalization?

\subsection{Experimental Setups}

\paragraph{Benchmarks.}
The experiments in this study are conducted using the Level-Based Foraging (LBF) \cite{papoudakis2020LBF} and the StarCraft Multi-Agent Challenge (SMAC) \cite{samvelyan2019SMAC} benchmarks. The former involves multiple agents collaborating in a 2D grid to collect scattered food items, where the sum of the agents' levels must exceed the food's level to successfully collect it and receive a reward. The latter is a complex real-time strategy game that requires multiple agents to cooperate to achieve objectives, such as destroying enemy units or protecting allied ones. SMAC provides a dynamic state space, heterogeneous unit types, and diverse map settings, effectively evaluating the agents' abilities in both cooperation and competition.

\paragraph{Baselines.}
To evaluate whether RGP performs well on these benchmark tasks as they continuously appear, we apply it to a popular value-based method, QMIX \cite{rashid2020QMIX}. Additionally, we use MACPro \cite{yuan2024MACPro}, EWC \cite{kirkpatrick2017EWC}, MAS \cite{aljundi2018MAS}, Replay, L2, and Finetuning as baselines, where MACPro is the latest method in Co-MARL. 

Furthermore, to validate the role of relation patterns in multi-agent cooperation, we combine the relation capturer with QMIX and compare it with common value-based and policy-based methods. Specifically, in SMAC, we use methods such as VDN \cite{sunehag2017VDN}, QMIX \cite{rashid2020QMIX}, WQMIX \cite{rashid2020WQMIX}, QTRAN \cite{son2019QTRAN}, ResQ \cite{shen2022ResQ}, HPN \cite{jianye2022HPN}, and FtQMIX \cite{hu2023ftqmix} as baselines. Additionally, we explore policy-based methods like MAA2C, COMA \cite{foerster2018COMA}, MADDPG \cite{lowe2017MADDPG}, MAPPO \cite{yu2022MAPPO}, and IPPO \cite{de2020IPPO} in LBF.

\subsection{Continual-tasks Performance and Ablation}

\paragraph{CL Performance Comparison.}
In Figure \ref{FIG:3}, we present a comprehensive result of the continual learning experiments conducted in SMAC, highlighting the superior performance of RPG. RPG outperforms other methods in both mitigating catastrophic forgetting and learning new knowledge. Specifically, the gray background indicates the current task and the training procedure synchronously tests all previous tasks using the current agent parameters. The results show that methods such as EWC, Replay, and MAS experience a performance drop after learning the first task, and fail to achieve optimal performance when learning new tasks, exhibiting weak capabilities in mitigating catastrophic forgetting and acquiring new knowledge. This indicates that directly applying CL methods to MARL settings is not feasible. Finetuning and L2 represent the upper bounds for plasticity and stability, respectively. However, Finetuning exhibits the worst catastrophic forgetting performance as it rapidly forgets knowledge after each task is learned, while L2 regularization, due to its excessive constraints, prevents the agent from adapting to task changes. 

The Co-MARL method MACPro, while performing well among the baselines, is inferior to RPG in both single-task performance and plasticity.
RPG adapts to 8m\_vs\_9m training faster than MACPro. Moreover, RPG's agent parameter size (112K) is smaller than MACPro's (377K), and MACPro's dynamic policy head further increases the parameter size.
Overall, RPG demonstrates comparable performance to L2 in ant-forgetting and is more adaptable to new tasks.
In addition, we compare RPG with several mentioned baselines on the continual tasks of LBF, as detailed in the Appendix.

\begin{table}
    \centering
    \begin{tabular}{ccccc}
        \toprule
        Methods  & 5m\_vs\_6m & 12m & 5m & 8m\_vs\_9m\\
        \midrule
        RPG     & \textbf{0.875}     & \textbf{1.00} & \textbf{1.00}     & 0.91 \\
        w/o \( \mathcal{L}_\text{rc} \)     & 0.86     & 1.00 & 1.00     & \textbf{0.95} \\
        w/o \( \mathcal{L}_\text{hy} \)  & 0.84     & 1.00 & 1.00     & 0.875 \\
        w/o \( \mathcal{L}_\text{rc},\mathcal{L}_\text{hy} \) & 0.865     & 1.00 & 1.00     & 0.86 \\
        RPG(EWC) & 0.875     & 1.00 & 1.00     & 0.93 \\
        RPG(MLP) & 0.84     & 1.00 & 1.00     & 0.59 \\
        \bottomrule
    \end{tabular}
    \caption{Ablation analysis of plasticity performance. ``w/o \(\mathcal{L}\)'' indicates the removal of \(\mathcal{L}\) from the RPG, and ``(\(\cdot\))'' denotes the use of \(\cdot\) as a replacement.}
    \label{tab:1}
\end{table}

\begin{figure}
    \centering
    \includegraphics[width=1.\columnwidth]{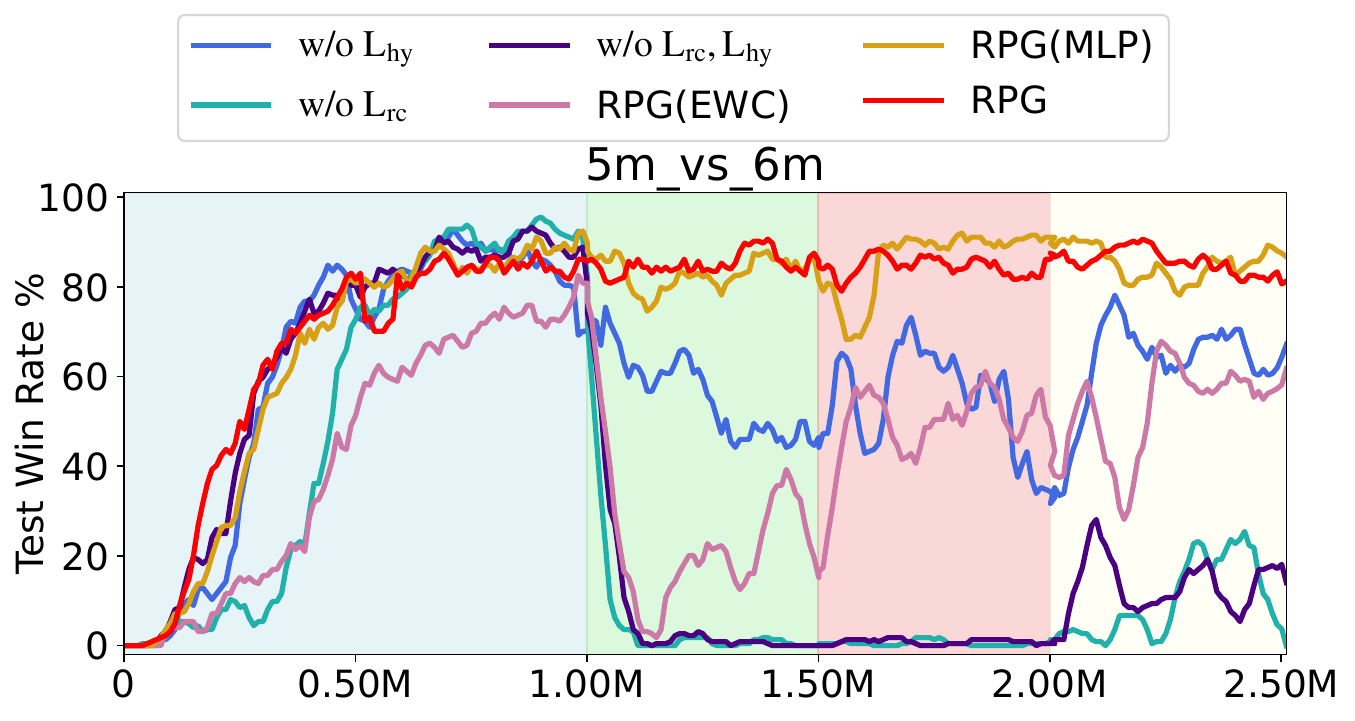}
    \caption{Ablation analysis of anti-forgetting performance. The 4 different colored backgrounds correspond to the training processes of \{5m\_vs\_6m, 12m, 5m, and 8m\_vs\_9m.\} Each curve represents the agent’s average win rate on 5m\_vs\_6m at distinct testing stages.}
    \label{FIG:4}
\end{figure}

\paragraph{Ablation Study.}
In this section, we investigate the impact of different components in RPG through an ablation study conducted on SMAC.
First, we remove the loss term \( \mathcal{L}_\text{rc} \) from the scalable relation capturer to obtain w/o \( \mathcal{L}_\text{rc} \), in order to study its effect on preventing catastrophic forgetting. Next, we replace \( \mathcal{L}_\text{rc} \) with EWC in RPG (denoted as RPG(EWC)) and apply it to the relation capturer to validate the effectiveness of our parameter importance estimation method. Then, in w/o \( \mathcal{L}_\text{hy} \), we eliminate the hypernetwork regularization loss term \( \mathcal{L}_\text{hy} \) to investigate whether a free hypernetwork could still facilitate continual learning. Additionally, we introduce RPG(MLP), which replaces the decision-maker generated by the hypernetwork with an MLP for decision-making. Finally, we remove both \( \mathcal{L}_\text{rc} \) and \( \mathcal{L}_\text{hy} \) from RPG, and we refer to this setup as w/o \( \mathcal{L}_\text{rc}, \mathcal{L}_\text{hy} \).

Continual learning aims to balance stability and plasticity, requiring joint analysis of Table 1 and Figure 4 for a comprehensive assessment.
In Figure \ref{FIG:4}, we test the algorithm on 5m\_vs\_6m at four stages during training to evaluate its forgetting resistance. Table \ref{tab:1} shows the algorithm's performance at different stages of task completion.
We find that w/o \( \mathcal{L}_\text{rc} \), lacking regularization, improves plasticity, but its performance rapidly degrades when the second task begins. This occurs because relaxing the relation capture constraints enhances the agent's learning capacity but compromises its memory retention. Similarly, RPG(EWC) shows a similar performance drop, confirming the reliability of our parameter importance estimation. Removing \( \mathcal{L}_\text{hy} \) allows w/o \( \mathcal{L}_\text{hy} \) to continue learning new tasks, but its forgetting resistance declines, highlighting the need for a constrained hypernetwork.
Notably, while RPG(MLP) maintains coordination on 5m\_vs\_6m, it loses some plasticity, emphasizing the necessity of task-specific decision-makers. 
This is because the MLP cannot isolate the decision-makers of different tasks.
Finally, w/o \( \mathcal{L}_\text{rc}, \mathcal{L}_\text{hy} \) loses the ability to resist forgetting.

\subsection{Single-task Performance}

\begin{figure}
    \centering
    \includegraphics[width=1.\columnwidth]{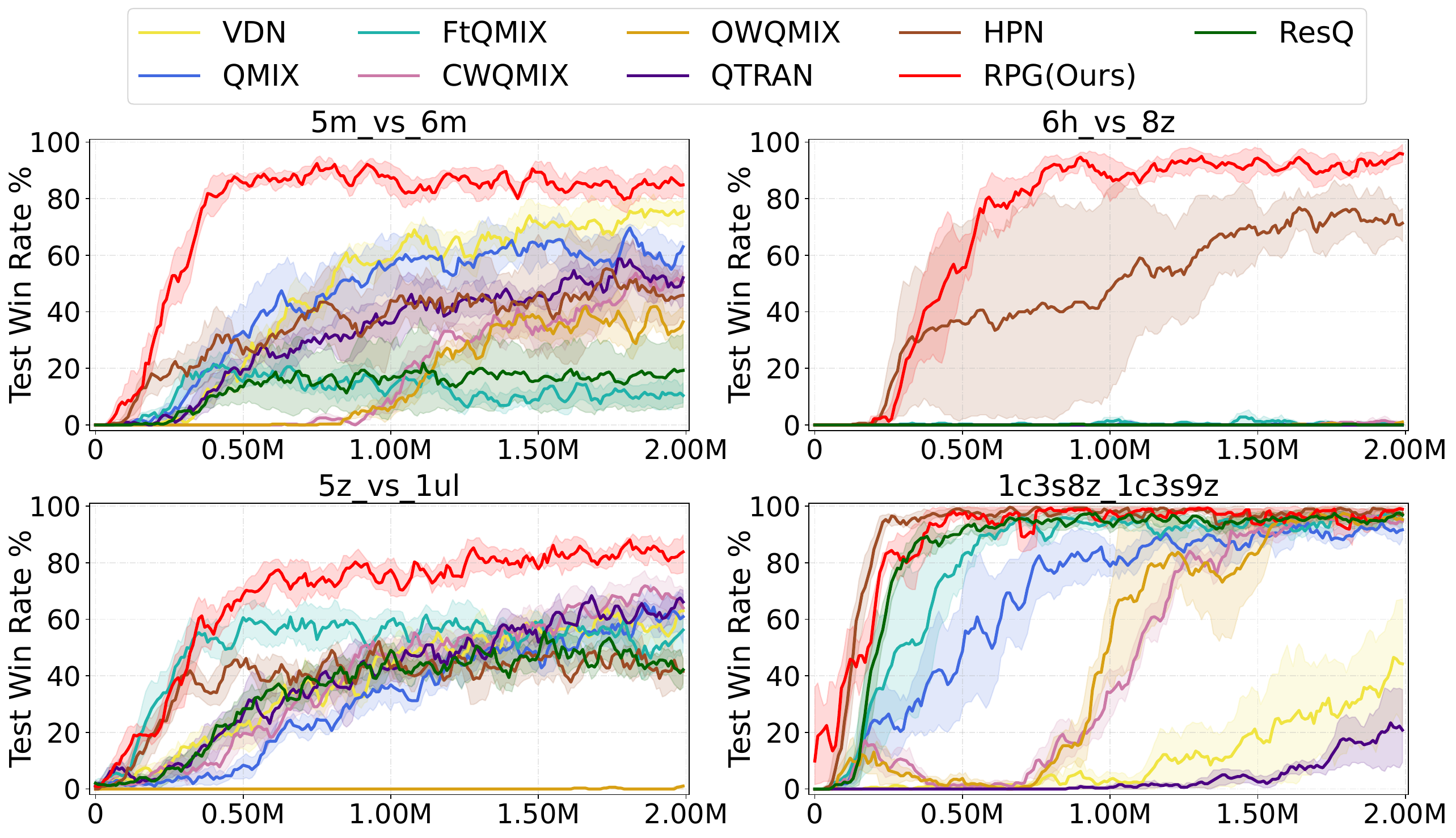}
    \caption{Algorithms Performance in 4 super-hard SMAC tasks.}
    \label{FIG:5}
\end{figure}

\begin{figure}
    \centering
    \includegraphics[width=1.\columnwidth]{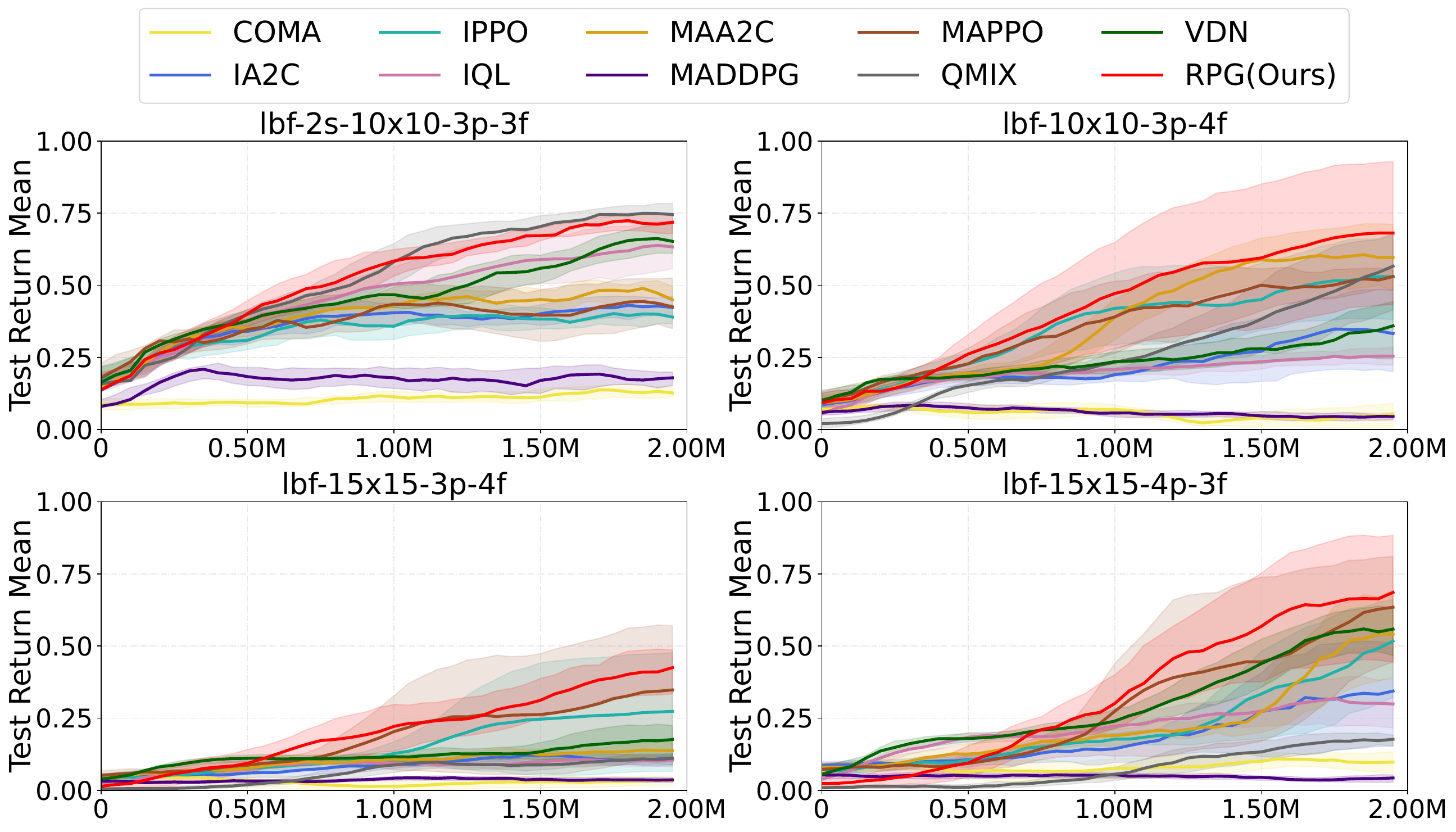}
    \caption{Algorithms Performance in 4 LBF tasks.}
    \label{FIG:6}
\end{figure}

In this section, we conduct comparative experiments to validate the coordination performance of our method in single tasks. Figure \ref{FIG:5} and \ref{FIG:6} show the comparison of RPG with baselines in SMAC and LBF.

In several super-hard maps of SMAC, RPG achieves the best performance, demonstrating that agents dynamically selecting interaction relationships with different teammates or opponents based on their own states is more conducive to achieving better coordination. In the LBF, particularly in maps with more entities and larger areas, RPG outperforms other baselines. This advantage is attributed to the ability of the relation pattern capturer to quickly identify key interactive entities in large-scale tasks, enabling efficient collaboration. These results indicate that the relation pattern is not only effective in complex, continual multi-agent cooperation tasks but also exhibits superior performance in more challenging single-task scenarios.

\begin{figure}
    \centering
    \includegraphics[width=1.\columnwidth]{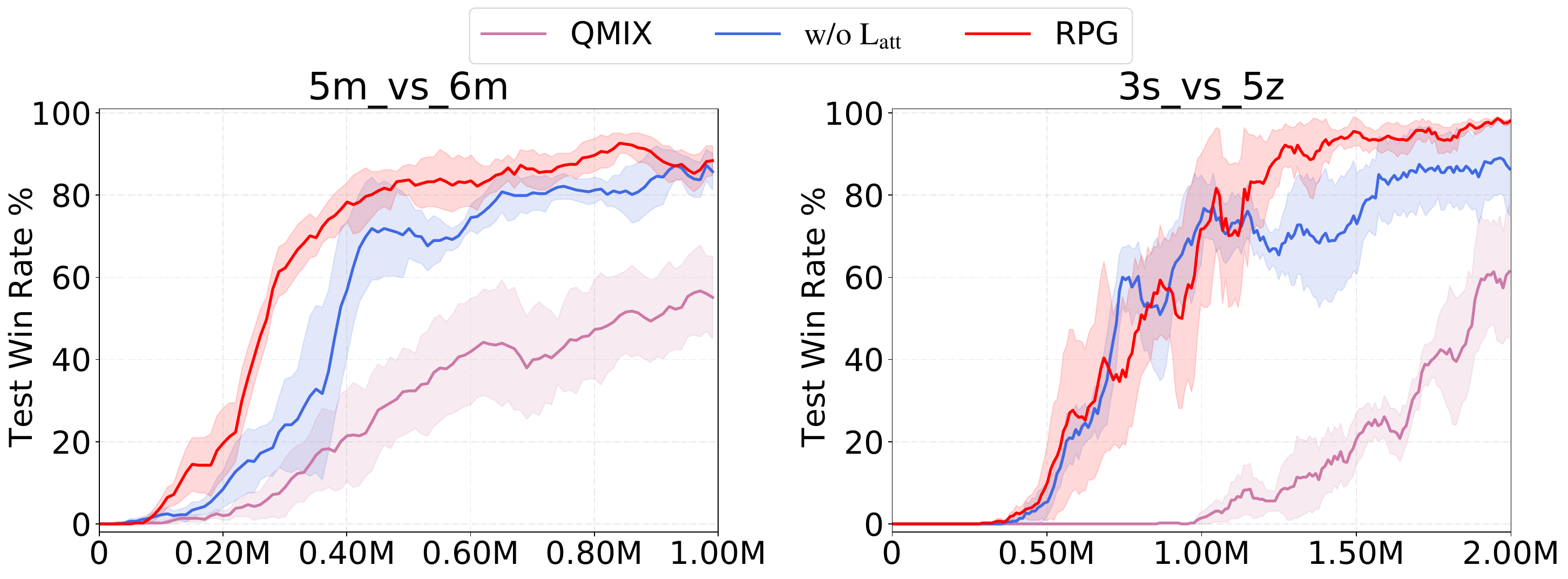}
    \caption{Ablation results of the sparse attention loss term.}
    \label{FIG:7}
\end{figure}

As shown in Figure \ref{FIG:7}, we conduct an ablation study on the sparse attention mechanism using w/o \( \mathcal{L}_\text{att} \) to analyze its impact on the generation of relational patterns by agents. The results indicate that removing sparse attention slightly reduces the performance of RPG, confirming that interacting with a few key entities facilitates better collaboration among agents.

\begin{table}
    \centering
    
    \begin{tabular}{cccc}
        \toprule
        Tasks  & RPG & EWC & MAS \\
        \midrule
        3m & $86.5\pm 8.3$ & $47.7\pm 13.9$ & $\bm{91.6\pm 8.5}$ \\
        8m & $\bm{100}$     & $95.3\pm 5.4$ & $99.6\pm 1.5$     \\
        10m  & $\bm{100}$     & $87.3\pm 7.9$ & $99.6\pm 1.5$     \\
        10m\_vs\_11m & $\bm{90.3\pm 7.9}$     & $89.8\pm 6.6$ & $63.1\pm 15.3$  \\
        15m & $\bm{96.3\pm 4.9}$     & $73.2\pm 10.6$ & $76.9\pm 16.6$  \\
        \bottomrule
    \end{tabular}
    \caption{Zero-shot results. Bold indicates the best performance.}
    \label{tab:2}
\end{table}

\subsection{Zero-shot Performance}

In addition to learning sequential tasks, agents should also learn to leverage prior coordination experience to handle unseen tasks. To end this, we compare RPG with EWC and MAS to assess its zero-shot generalization capability in tackling novel tasks. Specifically, in the SMAC scenarios, agents trained on continual tasks \{5m\_vs\_6m, 12m, 5m, 8m\_vs\_9m\} are tested on 5 additional unseen tasks to validate the generalization performance. As shown in Table \ref{tab:2}, compared to other baselines, RPG demonstrates superior generalization performance in 4 out of the 5 tasks. However, RPG's generalization ability on 3m does not match that of MAS. This indicates that in simple tasks with few agents, complex relation patterns may instead hinder simple cooperation. This highlights the ability of RPG to extract and retain generalizable relation patterns, enabling robust generalization to new tasks.

\section{Conclusion}
This paper focuses on the problem of continual coordination in MARL. To address this issue, we analyze the essence of continual multi-agent coordination—relation patterns—and propose a novel framework, RPG. RPG extracts general relation patterns from dynamic observation spaces and maps them to varying action spaces. Results demonstrate that RPG consistently outperforms baselines across various scenarios. However, RPG focuses solely on homogeneous multi-agent collaborative tasks and currently lacks the capability for perpetual knowledge acquisition. In future work, we aim to address these challenges. We hope RPG inspires further interest in continual multi-agent coordination research.

\section*{Acknowledgments}
This work was supported by the Fundamental Research Funds for the Central Universities (Grant No. 2024XKRC080) and the National Key Laboratory of Air-based Information Perception and Fusion (Grant No. 202300010M5001).


\bibliographystyle{named}
\bibliography{ijcai25}

\end{document}